\documentclass[aps,pra,twocolumn,superscriptaddress,nofootinbib,longbibliography,10pt]{revtex4-2}

\pdfoutput=1
\usepackage{microtype}
\usepackage{amsmath}
\usepackage{amssymb}
\usepackage{amsthm}
\usepackage{mathrsfs}
\usepackage{bm, dsfont}
\usepackage[table,usenames,dvipsnames]{xcolor}
\usepackage{booktabs}
\usepackage[normalem]{ulem}
\usepackage{physics}
\usepackage{graphicx}
\usepackage{diagbox}	
\usepackage{relsize, scalerel}
\usepackage[colorlinks=true,linkcolor=blue,citecolor=blue,urlcolor=blue]{hyperref}

\newcommand{\cH}{\mathcal{H}}

\newcommand{\be}{\begin{equation}}
\newcommand{\ee}{\end{equation}}
\newcommand{\id}{\mathds{1}}

\definecolor{EffBlue}{RGB}{30,90,200} 
\newcommand{\effcell}[2]{\cellcolor{EffBlue!#1!white}{#2}}

\newtheorem{definition}{Definition}

\begin{document}

\title{Certification of the genuine resolution of photon number resolving detectors}

\author{Jef Pauwels}
\thanks{These authors contributed equally.}
\affiliation{Department of Applied Physics, University of Geneva, CH-1205 Geneva, Switzerland}
\affiliation{Constructor University, 28759 Bremen, Germany}
\author{Towsif Taher}
\thanks{These authors contributed equally.}
\affiliation{Department of Applied Physics, University of Geneva, CH-1205 Geneva, Switzerland}
\author{Roope Uola}
\affiliation{Department of Physics and Astronomy, Uppsala University, Box 516, 751 20 Uppsala, Sweden}
\affiliation{Nordita, KTH Royal Institute of Technology and Stockholm University, Hannes Alfv\'ens v\"ag 12, 10691 Stockholm, Sweden}
\author{Boris Korzh}
\affiliation{Department of Applied Physics, University of Geneva, CH-1205 Geneva, Switzerland}
\author{Nicolas Brunner}
\affiliation{Department of Applied Physics, University of Geneva, CH-1205 Geneva, Switzerland}
\author{Pavel Sekatski}
\affiliation{Department of Applied Physics, University of Geneva, CH-1205 Geneva, Switzerland}

\begin{abstract}
Photon-number-resolving (PNR) detectors are essential components of photonic quantum technologies, yet thus far, no practical metric exists to certify how many photons they can genuinely resolve in a single measurement. Here we introduce an operational framework for quantifying the capability of a PNR detector to distinguish between different numbers of photons, i.e. its genuine resolution. In turn, we develop a practical and scalable protocol for certifying the genuine resolution of a detector, which is based on coherent state probes. We apply the method to a 28-pixel photon-number-resolving superconducting nanowire single-photon detector (PNR-SNSPD) and certify genuine four-outcome resolution. Our work highlights the critical requirements in terms of detector efficiency towards achieving high genuine resolution. This approach provides an operational benchmark for PNR detectors and fills a crucial gap in the characterization of photonic quantum devices.
\end{abstract}

\maketitle

\section{Introduction}

PNR detectors are devices that distinguish how many photons arrive in a given time window and they underpin a growing range of photonic quantum technologies. In the field of quantum communication, PNR detectors enable receiver error rates below the standard quantum limit~\cite{becerra2015}, improve heralded single-photon sources by rejecting multi-photon events~\cite{kaneda2019, davis2022}, certify the intended operation of quantum key distribution systems~\cite{moroder2009, dynes2018}, and can also be used for the direct generation of quantum random numbers~\cite{eaton2023}. In quantum sensing, PNR detectors enable enhanced phase estimation~\cite{afek2010, you2021} as well as surpass the Rayleigh limit in imaging applications~\cite{tenne2019}. In quantum computing, PNR detectors are required in universal linear-optics schemes~\cite{Knill2001, kok2007}, fusion-based models~\cite{Bartolucci2023, psiquantum2025manufacturable}, boson-sampling~\cite{Zhong2020,Madsen2022}, as well as for generating cat states~\cite{konno2024, winnel2024}. 

Over the past two decades, several PNR detector technologies have matured considerably, most notably transition-edge sensors (TES), which are bolometric detectors capable of measuring the total energy deposited~\cite{Lita2008, fukuda2011, morais2024precisely}. Superconducting nanowire single-photon detectors (SNSPD), which have outperformed TES in other metrics such as array size~\cite{oripov2023}; timing jitter~\cite{korzh2020}; count rates~\cite{craiciu2023, resta2023}; and minimum photon energy~\cite{taylor2023}, can also resolve the photon number by either operating in an array~\cite{Divochiy2008, cheng2023100, stasi2024enhanced, ding2025}, if the input is multiplexed~\cite{natarajan2013}, and most recently, by leveraging their intrinsic PNR capability~\cite{cahall2017, zhu2020resolving, los2024high}. These advances have resulted in a rush of commercialization efforts for both types of detectors~\cite{Madsen2022,stasi2024enhanced,los2024high, psiquantum2025manufacturable}. 

The rapid development and broad range applications of PNR detectors make their precise certification a task of utmost importance. Yet, the complexity of these devices renders their characterisation challenging. So far, several approaches have been investigated. Quantum measurement tomography~\cite{Luis1999, Fiurasek2001,Dariano2004} allows in principle for the full reconstruction of the quantum measurement operators (POVM) describing a PNR detector~\cite{Lundeen2009, humphreys2015}, while other methods~\cite{jonsson2019evaluating,provaznik2020benchmarking} have been proposed to benchmark their performance in specific tasks. Yet, from a practical perspective, these methods have significant drawbacks, as they require a prohibitively large number of probe states and/or complex setups. More importantly, we are still missing a practical figure of merit for quantifying the intrinsic capability of a PNR detector to deduce the number of photons that it received within a single measurement cycle, i.e. within a time window comparable to its temporal resolution. 

\begin{figure*}[t!]
    \centering
    \includegraphics[width=\textwidth]{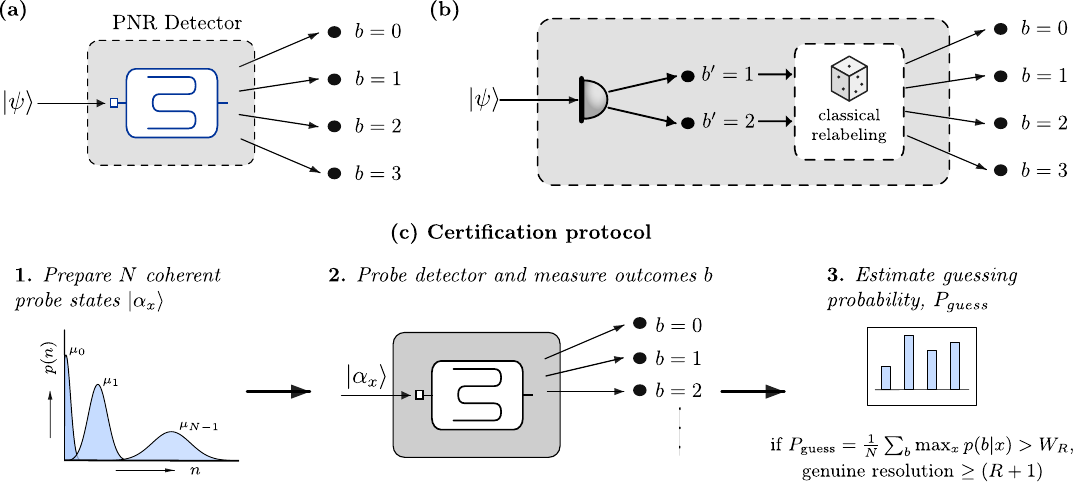}
    \caption{\textbf{Certifying the genuine resolution of a PNR detector.} (a) Our task is to quantify the ability of a given PNR detector to count the number of photons. Consider the device features four possible outcomes $b$. The main question is whether this device is indeed able to count the number of incoming photons, or whether it can be simulated by a simpler detector with lower resolution. As illustrated in (b), one can imagine a device featuring a simple binary (click/no-click) detector followed by classical post-processing to produce the observed four-valued outcome. (c) In order to assess the genuine resolution of our detector, we devise a simple, efficient and scalable test based on a guessing game. In each round, the user produces a coherent state with intensity $\mu_x$ (chosen according to a random classical input $x$) which is sent to the detector. From the measurement output $b$, the goal is to guess the classical input. The final score is given by the average guessing probability $P_{ \rm guess}$. Intuitively, the higher the resolution, the larger the score will be. More precisely, by comparing the score $P_{ \rm guess}$ with a witness $W_R$ we obtain lower bounds on the genuine resolution of the detector.    }
    \label{fig:concept}
\end{figure*}

To illustrate the importance of this point, imagine a vendor offering a PNR detector promising the ability to count up to 10 photons. How can a client verify this claim? More generally, how can one distinguish a detector performing a high-resolution quantum measurement, from a simple low-resolution device (e.g. a binary click/no-click detector) followed by some complex classical post-processing? This distinction between a genuinely high-resolution detector and a coarse detector followed by classical post-processing is illustrated in Fig.~\ref{fig:concept}a,b.

In this work, we introduce the notion of genuine resolution of a PNR detector, based on a precise and operational framework using concepts from the formal theory of quantum measurements~\cite{d2005classical,oszmaniec2017simulating,guerini2017operational}. Importantly, we show that achieving high genuine resolution requires high efficiency, by deriving lower bounds on the threshold efficiency of an (otherwise) ideal PNR detector. In turn, we present a simple and scalable certification protocol (Fig.~\ref{fig:concept}c) for obtaining directly relevant lower bounds on the genuine resolution. The protocol can be readily implemented in practice, as it only requires the preparation of coherent states in a prepare-and-measure scenario. 

We demonstrate the method experimentally on a 28-pixel PNR superconducting nanowire single-photon detector (PNR-SNSPD). In this case, we are able to certify a genuine resolution of 4, which operationally means that we exclude the possibility that our detector actually consists of a coarser device (three or less outcome measurement) followed by classical post-processing. This is possible here, thanks to a nominal single-photon detection efficiency of $\sim$85\% and a well-resolved readout scheme. Our approach requires only partial trust in the source calibration and bypasses the need for a full measurement tomography, thus providing an operational benchmark of PNR detectors.

%% ============================================================
%% RESULTS
%% ============================================================
\section{Results}

\subsection{Defining genuine resolution}

Consider a device that can supposedly resolve high photon numbers, hence capable of counting the number of incoming photons in a single shot pulse, say from 0 up to $K-1$ photons. This device is thus expected to produce (at least) $K$ distinct outcomes. Yet, the mere existence of $K$ output labels does by no means imply that the device has the ability to count the number of photons, or more generally resolve $K$ genuinely distinct (orthogonal) measurement outcomes. Indeed, imagine a detector that internally performs only a binary measurement---distinguishing, say, vacuum from one or more photons---and then randomly reassigns this binary result across $K$ output labels. Clearly, such device can in principle produce any number of outcomes $K$, yet its resolution is still only binary. 

It is therefore important to define the resolution of a measurement independently of its number of outcomes. Here we propose a definition of resolution at an operational level, building upon recent advances in the theory of quantum measurements, investigating the possibility of simulating certain measurements via others \cite{Uola2023,oszmaniec2017simulating,guerini2017operational,ioannou2022simulability}. Specifically, a measurement is said to be $R$-outcome simulatable~\cite{guerini2017operational} if it is indistinguishable from a detector built out of measurements with
(at most) $R$ outcomes, followed by classical post-processing. In turn, we say that a measurement that is provably not $R-$outcome simulatable has a genuine resolution of (at least) $R+1$ (see Definition~\ref{def:genuine} in Methods~\ref{sec:methods-formal}). This represents a device-agnostic, operational notion of resolution: it characterizes what the detector can intrinsically resolve, irrespective of how many outcome labels it assigns.

\begin{table*}[t!]
\centering
\begin{tabular}{|c|c|c|c|c|c|c|c| }
\hline
\diagbox{${\mathcal{H}}_m$}{ $R$}  & 3 & 4 & 5 & 6 & 7 &8& 9 \\
\hline
$2$& \effcell{38}{61.8} & -&  -&- &- &- &-  \\
$3$ & \effcell{30}{50.0}& \effcell{53}{81.1} & - & -&- &-&-  \\
$4$ &\effcell{22}{39.8} & \effcell{48}{73.8} &  \effcell{59}{88.8} & - & -& -&- \\
$5$ & \effcell{17}{34.0} & \effcell{42}{67.4} & \effcell{55}{83.4}&  \effcell{63}{92.7} & -& -&- \\
$6$ & \effcell{13}{29.2} & \effcell{38}{61.9}&  \effcell{52}{79.8} & \effcell{59}{88.5} & \effcell{65}{94.8} & -&- \\
$7$ & \effcell{10}{25.8} & \effcell{35}{57.5}&  \effcell{49}{75.9}& \effcell{57}{86.0} & \effcell{62}{91.6} & \effcell{66}{96.2}&- \\
$8$ &\effcell{8}{23.0} & \effcell{32}{53.6}&  \effcell{47}{72.6}& \effcell{55}{83.4} & \effcell{60}{89.5} & \effcell{64}{93.6}& \effcell{67}{97.1}\\
\hline
\end{tabular}
\caption{\textbf{Threshold efficiencies for the efficiency-limited detector.} The threshold efficiency $\eta_{\mathrm{th}}$ (in \%) above which the efficiency-limited photon-number measurement in Eq.~\eqref{eq:eff-lim} has genuine resolution $R$ in the subspace $\cH_m$, i.e. with up to $m$ photons. %photon numbers $[0,1,\ldots,m]$. 
Darker blue cells indicate larger threshold efficiencies. Observe that $\eta_{\rm th}$ increases with $R$, but decreases with $m$.  Note that the case $R=2$ is omitted, since for any nonzero efficiency 
$\eta_{\rm th}>0$ the measurement has genuine two-outcome resolution, on any subspace ${\cH}_m$ (with $m \geq 1$).}
\label{tab:eff-lim}
\end{table*}

In order to capture the ability of a PNR detector to resolve a single-shot photon number, one aims to determine its resolution in the \emph{low photon-number regime}. Indeed, existing detectors can distinguish many intensity levels of bright light e.g. modern smartphone cameras already achieve sub-shot-noise intensity resolution~\cite{Sanguinetti2014}. What is truly challenging, and essential for applications in quantum technologies, is resolving different photon numbers in the regime close to the single-photon level. We capture this through the notion of \emph{genuine subspace resolution}. We say that a detector has genuine resolution of at least $R+1$ in the subspace $\cH_m = \mathrm{span}\{\ket{0},\ldots,\ket{m}\}$ if the measurement restricted to that subspace---obtained by projecting its operators onto ${\cH}_m$ ---is not $R$-outcome simulatable (Definition~\ref{def:subspace} in Methods~\ref{sec:methods-formal}).

\medskip

To build intuition, consider an ideal photon-number detector preceded by a lossy channel with transmission (efficiency) $\eta$. In the subspace ${\cH}_m$, this ``efficiency-limited'' model is described by the measurement operators (see Methods~\ref{sec:methods-eff-lim})
\be\label{eq:eff-lim}
M_b^{(\eta,m)} = \sum_{ n = b}^m \eta^b (1-\eta)^{n-b}\binom{n}{b}\ketbra{n},
\ee
where $\ket{n}$ denotes the Fock state with $n$ photons and $b$ denotes the measurement outcome, with normalisation given by $\sum_b M_b^{(\eta,m)} = \mathds{1}_m$. 

Even if this detector has a very large number of outcomes, its genuine resolution strongly depends on $\eta$. As an illustration, consider the simplest case of $m=2$, i.e. restricting to the subspace $\cH_2$ with 0, 1 and 2 photons. For $\eta=1$, the measurement has clearly maximal resolution $R=3$. But when $\eta \leq \eta_{\rm th} = \frac{\sqrt{5}-1}{2} \approx 61.8\%$ (the inverse golden ratio), the detector becomes $2$-simulatable, i.e. it can be perfectly simulated via binary  measurements assisted by classical post-processing. This shows that detectors featuring a high genuine resolution must have a high enough efficiency.

In general, the threshold efficiency $\eta_{\rm th}$ depends on the subspace dimension $m$ and the genuine resolution $R$, and can be estimated via linear programming (see Methods \ref{sec:methods-eff-lim}). In Table~\ref{tab:eff-lim} we present the results for $m \leq 8$. We observe that  higher resolution generally requires higher efficiencies, whereas increasing the subspace dimension allows for lower efficiencies (for fixed resolution $R$). Clearly, the certification of maximal resolution $R=m+1$ becomes very demanding in terms of efficiency as the subspace dimension $m$ grows. For example, certifying resolution $R=5$ in the subspace ${\cH}_4$ already requires $\eta > 88.8\%$. Hence, loss severely limits the genuine resolution of PNR detectors. This reveals a fundamental scaling challenge for photon-number resolution: the threshold efficiency required to achieve resolution $R$ in a subspace of dimension $m+1$ is governed not only by the photon number scale but by the ratio $R/(m+1)$. 
When $R\ll m$, the detector is required to resolve only $R$ distinguishable clusters over a broader photon-number range. In contrast, when $R\approx m+1$, it must resolve nearly every photon-number level individually, leading to a substantially higher efficiency threshold. This challenge becomes progressively more severe at larger photon numbers.
The physical origin of this difficulty can be understood from the sensitivity of higher Fock states to optical loss: the distinguishability of $\ket n$ and $\ket {n+1}$ after a channel of transmitivity $\eta = 1- \varepsilon$ decreases as $1-n\, \varepsilon +O(\varepsilon^2)$, making resolution of consecutive photon numbers progressively more demanding at large $n$.

\subsection{A witness for certifying genuine resolution}
\label{sec:witnesses}

\begin{figure*}[t!]
    \centering
    \includegraphics[width=\textwidth]{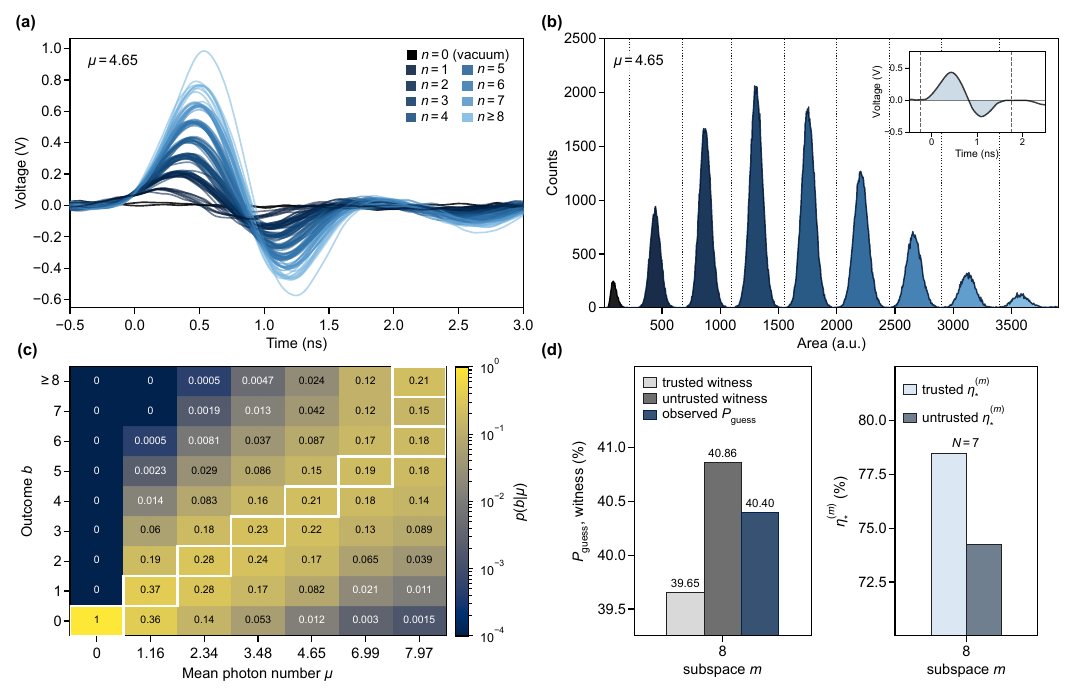}
    \caption{{\bf Data acquisition and analysis for the 28-pixel PNR-SNSPD.} \textbf{a)} Representative oscilloscope waveforms for an input coherent state ($\mu = 4.65$), where voltage corresponds to the number of simultaneous pixel clicks ($n$). \textbf{b)} Histogram of the integrated pulse areas for $\mu = 4.65$ for $2.5 \times 10^5$ non-zero detection events. Well-resolved peaks enable the unambiguous assignment of the (discrete) outcome $b \in \{ 0,...,8\} $ via the classification thresholds (dashed vertical lines). Inset: a single waveform on an expanded time axis. The shaded region indicates the 2~ns integration window. \textbf{c)} Measured conditional outcome distributions $p(b|\mu_x)$ for the input states $\mu_x$ with $x \ \{ 1,...,7\}$. For each output $b$, the most likely input $\mu_x$ is highlighted with a white frame. \textbf{d)} From the Table (c), the score $P_{ \rm guess}$ is computed and compared to the corresponding witness bounds. Here we certify, in the trusted model, a genuine resolution of $R=4$ in the subspace of up to $m=8$ photons (left). Additionally, we certify an effective efficiency (see main text) of approximately $77 \%$, compatible with the available detector parameters.
    }
    \label{fig: experiment data analysis}
\end{figure*}

The next natural question is how to test the genuine resolution of a PNR detector in practice. So far, we have considered the situation where a complete description of the measurement operators is available. For a realistic device, this information can in principle be obtained via full measurement tomography \cite{Lundeen2009,humphreys2015}, which requires a large number of fully characterized probe states; $O(m^2)$ for the subspace ${\cH}_m$. However, full measurement tomography is not necessary, and we will now present a far more efficient and economical approach for certifying genuine resolution. This is based on a simple guessing game in a prepare-and-measure setup, which requires only $O(m)$ probe states.

\medskip

As a warm-up, consider a guessing game, where a first party (Alice) receives a random input $x \in \{1,\dots,N\}$, encodes this input in a physical system, and sends it to another party (Bob). Upon receiving the system, Bob will make a measurement, producing outcome $b$. The goal for Bob is to infer Alice's input $x$. Accordingly, he guesses the value of $x$ that is most probable given the observed outcome $b$. The final score is given by the average probability of Bob guessing correctly 
\be\label{eq:Pguess}
P_{\rm guess} = \sum_b p(b) \max_x p(x|b)= \frac{1}{N}\sum_b \max_x\, p(b|x).
\ee
Clearly, in order to maximize the score, Bob should use a measurement that can distinguish well the different states of the physical system sent by Alice, i.e. a measurement with high genuine resolution. More formally, one can show that
\be\label{eq:source-agnostic}
P_{\rm guess} \leq \frac{R}{N} \,,
\ee
which follows from the fact that a measurement with at most $R$ distinct outcomes can identify at most $R$ states perfectly, together with the monotonicity of the guessing probability under classical post-processing and convex mixtures. Hence this game provides a witness for certifying a lower bound on genuine resolution. Indeed, when the observed data leads to a violation of the above inequality, one can certify a genuine resolution of (at least) $R+1$. The corresponding guessing-game protocol is sketched in Fig.~\ref{fig:concept}c.

So far, we have made no assumption on the physical system prepared by Alice, hence the model is source-independent. The best strategy for Alice would be to encode her input $x$ in orthogonal states, such as Fock states of different photon number, which can then be distinguished perfectly by a high-resolution detector. In practice, however, the preparation of Fock states is extremely challenging, which motivates us to consider an alternative scenario, where Alice encodes $x$ in (weak) coherent states of different intensities $\mu_x$. Since we want to certify resolution within a given photon number subspace, the intensities $\mu_x$ should be chosen in order to have large support in that subspace, i.e. the intensities cannot be too high. As a rule of thumb, we use $0 \leq \mu_x \leq m$, with the values $\mu_x$ spread over the interval. We will consider two different models, with varying level of trust in the source. First, in the ``trusted'' scenario we assume  a well-calibrated source and hence that all intensities $\mu_x$ are known. Second, we also consider an ``untrusted'' scenario, in the spirit of semi-device-independent protocols \cite{VanHimbeeck2017,brask2026}. Here the source is partially characterized, in the sense that we assume only an upper bound on the intensities $\mu_x \leq I$. 

The next step is now to adapt the witness in Eq. \eqref{eq:source-agnostic} to this alternative scenario. The main difference comes from the fact that the weak coherent states have a significant overlap and hence cannot be perfectly distinguished from each other, even by an ideal PNR detector. To take this key point into account, we must now compute refined upper-bounds on the witness \eqref{eq:source-agnostic}. To do so, we observe that when considering a PNR detector which is not phase-locked to the source, its task consists in distinguishing phase-averaged coherent states, i.e. mixtures of Fock states weighted by the Poisson distribution
\be\label{eq:Poisson}
q(n|\mu_x) = \frac{\mu_x^n}{n!}\,e^{-\mu_x},
\ee
with mean photon number $\mu_x$. Of course, these distributions also extend out of the subspace $\cH_m$, as they feature terms with $n>m$ photons. In order to be conservative, we consider that the detector is ideal out of the subspace $\cH_m$, i.e. it can perfectly distinguish Fock states with $n>m$ photons --note that among all detectors with the same subspace resolution $R$, this is the one achieving the highest guessing probability. The exact form of the witnesses and bounds are given in the Methods~\ref{sec:methods-witnesses}, for both the trusted and untrusted scenarios.

\subsection{Experimental demonstration}
\label{sec:experiment}

\begin{figure*}[t!]
    \centering
    \includegraphics[width=\textwidth]{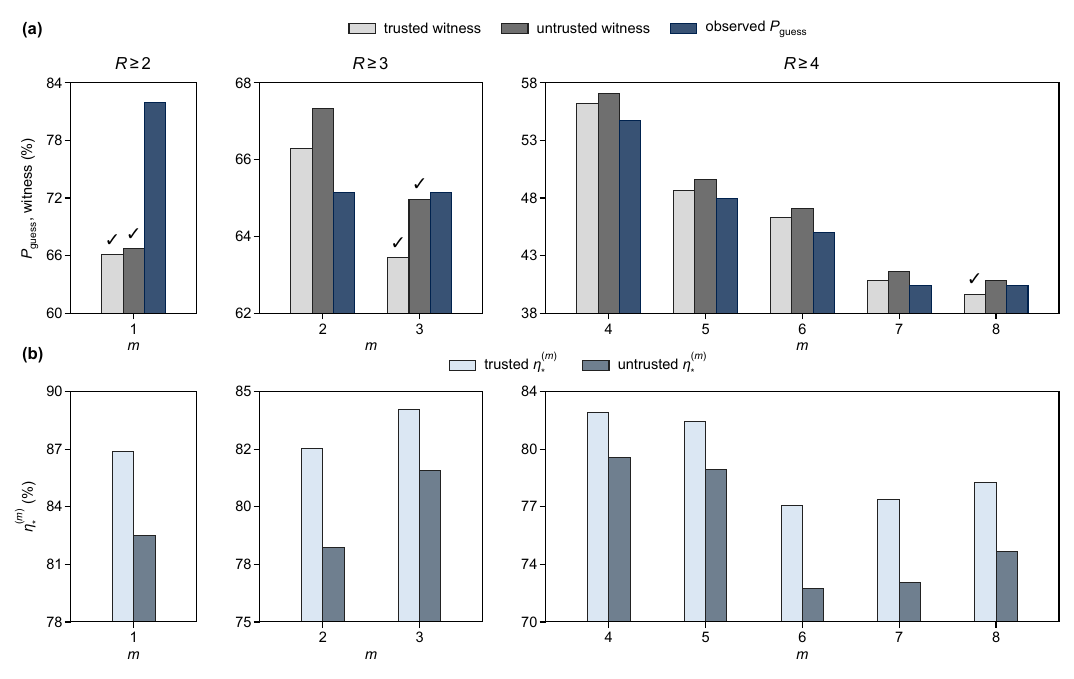}
    \caption{ 
    {\bf Experimental certification of genuine resolution and effective efficiency.} By considering various subsets of input states, the data in Fig.2(c) is used to certify lower bounds on the genuine resolution $R$ and the effective efficiency, for different photon number subspace $m$. (a) In the 0-1 photon subspace (m=1), a maximal genuine resolution of $R=2$ is certified. A genuine resolution $R \geq 3$ is certified in the subspace with up to $m=3$ photons; note that, with our data, this is not possible for $m=2$. Finally, in order to certify a resolution $R \geq 4$, we must increase the subspace up to $m=8$ photons. The statistical error is negligible in the trusted scenario, but increases in the untrusted model. (b) Additionally, for each case we estimate the effective efficiency. For low subspaces, i.e. $m\leq5$ we obtain values in the range of $82 \%$-$85 \%$, which drop to $75\%$ for higher subspaces. All numerical values are given in the Methods \ref{sec:methods-numerics}; see Tables~\ref{tab:certification} and \ref{tab:certified-efficiencies}.
    }
    \label{fig:experiment-certification}
\end{figure*}

We test our certification method using a PNR-SNSPD comprising of 28 interleaved active pixels, which are multiplexed onto a single readout channel in order to create signals of different amplitudes when different numbers of photons hit the detector simultaneously~\cite{stasi2024enhanced}. This detector is capable of producing up to $K=8$ outputs, i.e., 8 distinct voltage levels, as shown in Fig.~\ref{fig: experiment data analysis}a. The nominal single-photon detection efficiency is approximately 85\% (see Methods~\ref{sec:methods-setup} for details of the experimental setup and analysis). Figure~\ref{fig: experiment data analysis} outlines the experimental data acquisition and analysis workflow. 

We probe the genuine resolution of our detector in different subspaces, with up to $m=8$ photons, which is chosen to match the maximum number of outputs our detector is capable of. To do so, we implement the certification protocol using $N=7$ different coherent probe states, with intensities (mean-photon number per pulse) given by $\{\mu_i\}_{i=1}^7 = \{0, 1.16, 2.34, 3.48, 4.65, 6.99, 7.97\}$. For each input state, we record oscilloscope traces (Fig.~\ref{fig: experiment data analysis}a) corresponding to $2.5\times 10^5$ non-zero detection events, and we compute the time-integrated pulse area over a 2~ns window for each trace. The resulting distribution of integrated pulse areas (Fig.~\ref{fig: experiment data analysis}b) exhibits well-resolved peaks corresponding to distinct numbers of simultaneously triggered detector elements. By defining classification thresholds between adjacent peaks, each waveform is assigned to an outcome $b \in \{1,...,m\}$. The number of zero-photon outcomes, and subsequently the total number of outcomes, is estimated by analyzing the timestamps of waveforms from non-zero detection events and the total data acquisition time window (see Methods~\ref{sec:methods-setup}). From the data, we can estimate the conditional probability distribution $p(b|\mu_x)$ (Fig.~\ref{fig: experiment data analysis}c), from which we finally extract the average guessing probability $P_{\mathrm{guess}}$. The resulting genuine-resolution certificates are summarized in Fig.~\ref{fig:experiment-certification}a.

To start, let us consider the simplest case of $m=1$, i.e. restricting on the subspace of 0 and 1 photon, with the goal of certifying a resolution $R=2$. Considering the $N=2$ probe states of lowest intensity, we obtain $P_{\rm guess} = 81.90\%$. The witness bounds are computed as follows. For a trivial detector ($R=1$), the guessing probability is $50\%$ in the subspace $m=1$, but becomes unity outside of the subspace. Including this tail correction, we get a trusted bound of $66.10\%$. For the untrusted scenario, here and below, we choose a conservative estimate of the maximal intensity by including a $3\%$ calibration error i.e. $I=1.03\mu_1$, the bound is $66.73\%$. Therefore, our data certifies a genuine resolution $R=2$, with statistical significance still exceeding $300\sigma$ (see Methods~\ref{sec:methods-stats}).

Next, we aim to certify a genuine resolution $R=3$. With our detector, this becomes possible in the subspace $m=3$. Using $N=3$ input states (of lowest intensity), we obtain $P_{\rm guess} = 65.16\%$. This value exceeds the bounds for $R=2$ for both the trusted ($63.45\%$) and untrusted ($64.97\%$, with $I=1.03\mu_3$) scenarios. For the untrusted scenario this gives a significance of $\sim 3.4\sigma$.

In turn, to certify resolution $R=4$, we must consider a subspace with up to $m=8$ photons. Using all $N=7$ probe states, we get $P_{\rm guess} = 40.40\%$, exceeding the bound for $R=3$ for the trusted scenario ($39.7\%$) with a statistical significance of $\sim 21\sigma$. For untrusted scenario, the corresponding bound is $40.86\%$, so the observed estimate does not certify $R=4$ in this model. 

\subsection{Effective efficiency}

An additional benchmark that can be extracted from our protocol is a notion of ``effective efficiency'' for the PNR detector. The idea is to compare its performance (as quantified by the score $P_{\rm guess}$) to that of an (ideal) efficiency-limited detector~\eqref{eq:eff-lim}. Specifically, for each subspace $m$, one can compute the efficiency $\eta^*_m$ required for the efficiency-limited detector~\eqref{eq:eff-lim} to reproduce the observed value $P_{\rm guess}$ (see Methods~\ref{sec:methods-efficiency}), for both the trusted and untrusted scenario, see Fig.~\ref{fig:experiment-certification}b. Note that, in comparison to the resolution $R$ the obtained values of $\eta^*_m$ are significantly higher than the the threshold efficiencies $\eta_{\rm th}$ in Table.~\ref{tab:eff-lim}, this is due to the fact that our detector is not only limited by its efficiency, as shall be explained below. From our data, we can certify values of \(\eta^*_m\) in the range \(82\%\) to \(85\%\) for $m\leq 5$, dropping to about \(75\%\) for higher subspaces $m$. This is consistent with the nominal single-photon detection efficiency of approximately \(85\%\) and the expected combinatorial losses of the multiplexed architecture.

A useful feature of this benchmark is that $\eta^*_m$ subsumes all device-specific imperfections into a single, subspace-resolved figure of merit. Real PNR architectures are affected by a variety of error mechanisms beyond simple optical loss. In multiplexed detectors such as the PNR-SNSPD studied here~\cite{stasi2024enhanced}, combinatorial ambiguity -- arising when multiple photons impinge on the same pixel -- introduces resolution degradation that grows with photon number and is not captured by a global efficiency figure. In intrinsic energy-resolving devices such as transition-edge sensors, the limiting factor is instead the assignment of analog signals to discrete photon-number bins. Two detectors with identical nominal single-photon efficiencies can therefore exhibit markedly different resolving power in a given subspace. Extracting $\eta^*_m$ within each subspace of interest from a handful of coherent-state measurements directly benchmarks all these combined effects, and enables a direct comparison between different detector architectures. 

\section{Discussion}
 
We have defined an operational and model-agnostic notion of genuine resolution for PNR detectors. In turn, we have developed efficient and practical methods for certifying lower bounds on the genuine resolution in a specified low photon-number subspace. Implementing these methods experimentally using a 28-pixel PNR-SNSPD array, we report the certification of genuine four-outcome resolution in a subspace with up to eight photons. 

Our methods avoid full detector tomography and are based only on directly measurable input-output statistics via weak coherent probes. As such, they provide a scalable route for benchmarking PNR detectors across different architectures and comparing their performance. To facilitate the implementation of this certification protocol, we provide accompanying source-code and an interactive and ready-to-use online application for evaluating the resolution witness. 

\section{Methods}
\label{sec:methods}

\subsection{Definition of resolution}
\label{sec:methods-formal}

In order to provide an operational definition for the resolution of a measurement, we take advantage of concepts and tools developed in the context of quantum measurement theory; see e.g. \cite{Uola2023}. A quantum measurement with $K$ outcomes is described by a positive operator-valued measure (POVM): a set of positive semidefinite operators $\{M_b\}_{b=1}^K$ satisfying $\sum_b M_b = \id$, where $\id$ is the identity. The probability of outcome $b$ given input state $\rho$ is given by the Born rule $p(b|\rho) = \tr(M_b\, \rho)$. 

As argued in the main text, it is important to distinguish the resolution $R$ of a measurement from the number of outcomes $K$, as a $K$-outcome can be simulated with coarser measurements, i.e. with $R<K$. Formally, we can define the notion of a genuine resolution of a measurement, which captures the lowest possible resolution required for simulating a given POVM.

\begin{definition}[Genuine resolution]
\label{def:genuine}
A POVM $\{M_b\}_{b=1}^K$ has genuine resolution of (at least) $R+1$ if it cannot be simulated by measurements with at most $R$ outcomes. That is, there exist no collection of $R$-outcome POVMs $\{E_{i|\lambda}\}_{i=1}^R$ and probabilities $p(\lambda)$, $p(b|i,\lambda)$ such that
\begin{equation}\label{eq:simulability}
M_b = \sum_\lambda \sum_{i=1}^R p(\lambda)\, p(b|i,\lambda)\, E_{i|\lambda} \quad \text{for all } b.
\end{equation}
The exact genuine resolution is therefore the largest value $R+1$ for which no $R$-outcome simulation is possible.
\end{definition}

As argued in the main text, for PNR detectors, it is important to further refine this definition in order to capture resolution in a subspace, of typically low photon-number.

\begin{definition}[Genuine subspace resolution]
\label{def:subspace}
A POVM $\{M_b\}_{b=1}^K$ has genuine resolution at least $R+1$ on the subspace ${\cH}$ if the restricted POVM $\{\bar{M}_b := \Pi_{{\cH_m}}\, M_b\, \Pi_{{\cH_m}}\}_{b=1}^K$ has genuine resolution at least $R+1$, where $\Pi_{{\cH}}$ is the projector onto $\cH$.
\end{definition}

For PNR detectors, it is natural to consider the subspace ${\cH}_m$ spanned by Fock states with up to $m$ photons. Note that $R$-simulatability of a given POVM can be tested with linear programming, see next section.

\subsection{Efficiency-limited detector}
\label{sec:methods-eff-lim}

The efficiency-limited photon-number measurement is modeled by a projective measurement in the Fock basis preceded by a loss channel $\Lambda_\eta$ with transmission $\eta$, yielding the POVM elements in Eq.~\eqref{eq:eff-lim}. Restricted to the subspace ${\cH}_m = \mathrm{span}\{\ket{0},\ldots,\ket{m}\}$, each element is diagonal in the Fock basis and represented by the column-stochastic matrix 
$\mathrm{M}_{bn} = \bra{n}\bar{M}_b\ket{n}$.

For input states diagonal in the Fock basis, all measurements in the decomposition~\eqref{eq:simulability} can be taken diagonal without loss of generality. Furthermore, it suffices to consider extreme diagonal $R$-outcome POVMs, which correspond to deterministic partitions of the Fock states $\{\ket{0},\ldots,\ket{m}\}$ into $R$ disjoint subsets. The number of such partitions is the Stirling number of the second kind $S(m+1,R)$.

$R$-outcome simulatability can then be tested via the following linear feasibility problem. Denoting each partition $\mathcal{P}_\lambda = \{\mathds{S}_{i|\lambda}\}_{i=1}^R$, find non-negative variables $p(b,\lambda|i)$ satisfying
\begin{align}\label{eq:LP}
\mathrm{M}_{bn} &= \sum_{\mathcal{P}_\lambda}\sum_{i=1}^R p(b,\lambda|i)\,\mathds{1}[n \in \mathds{S}_{i|\lambda}]
\quad \forall\, b,n, \nonumber\\
\sum_{b,\lambda} p(b,\lambda|i) &= 1 \quad \forall\, i, \nonumber\\
\sum_b p(b,\lambda|i) &= \sum_b p(b,\lambda|i') \quad \forall\, \lambda, i, i',
\end{align}
wherer $\mathds{1}[...]$ is the indicator function. Note that the last constraint ensures that $\lambda$ is independent of $i$, enabling the decomposition $p(b,\lambda|i) = p(b|i,\lambda)\,p(\lambda)$.

We solve this linear program while varying $\eta$ (in steps of 0.1\%) to find the threshold efficiency $\eta_{\rm th}$ below which the restricted measurement becomes $R$-outcome simulable. The results are reported in Table~\ref{tab:eff-lim}. For the three-outcome case ($m=2$, $R=2$), the value $\eta_{\rm th} = (\sqrt{5}-1)/2$ can be obtained analytically, by constructing an explicit decomposition as a mixture of the three extremal binary POVMs $\{|0\rangle\!\langle 0|, |1\rangle\!\langle 1|+|2\rangle\!\langle 2|\}$, $\{|0\rangle\!\langle 0|+|1\rangle\!\langle 1|, |2\rangle\!\langle 2|\}$, and $\{|0\rangle\!\langle 0|+|2\rangle\!\langle 2|, |1\rangle\!\langle 1|\}$, chosen with probabilities $(2\eta_{\rm th} (1-\eta_{\rm th} ),\, {\eta_{\rm th}}^2,\, (1-\eta_{\rm th})^2) $.

\subsection{Resolution witness}
\label{sec:methods-witnesses}

In a prepare-and-measure scenario with $N$ uniformly sampled input states $\{\rho_x\}$ and measurement $\{M_b\}$, the guessing probability is $P_{\rm guess} = \frac{1}{N}\sum_b \max_x \tr(M_b\,\rho_x)$.
If $\{M_b\}$ is $R$-outcome simulable, then we get $P_{\rm guess} \leq R/N$ as in Eq.~\eqref{eq:source-agnostic}.

For our PNR experiment, the source prepares $N$ coherent states $\ket{\alpha_x}$ with mean photon numbers $\mu_x = |\alpha_x|^2$. Since the detector is not phase-locked to the source, the effective input states are phase-averaged,
$
\rho_x = \sum_{n=0}^\infty q(n|\mu_x)\,\ketbra{n},
$
with $q(n|\mu)$ the Poisson distribution~\eqref{eq:Poisson}. For such diagonal states, the optimal measurement is also diagonal in the Fock basis.

To derive subspace-resolution witnesses, consider a measurement $\{M_b\}$ whose restriction to ${\cH}_m = \mathrm{span}\{\ket{0},\ldots,\ket{m}\}$ has resolution at most $R$. Since the states are diagonal, the measurement elements decompose as $M_b = \bar{M}_b \oplus M_b^\perp$, where $\bar{M}_b$ denotes the restriction to ${\cH}_m$. Any such measurement is a post-processing of the finer POVM $\{M_{b,0} = \bar{M}_b,\; M_{b,1} = M_b^\perp\}$, so the guessing probability satisfies
\begin{equation}\label{eq:WRm-methods}
    P_{\rm guess} \leq \frac{1}{N}\Big(\sum_b \max_x \tr(\bar{M}_b\,\rho_x)   +\sum_b \max_x \tr(M_b^\perp\,\rho_x)\Big).
\end{equation}

Since we assume that the POVM $\{M_b^\perp\}$ outside the subspace has perfect resolution, and the states are diagonal, we get
\[
\sum_b \max_x \tr(M_b^\perp\,\rho_x) \leq \sum_{n \geq m+1} \max_x q(n|\mu_x).
\]
When $\max_x \mu_x \leq m+1$, we have that $\max_x q(n|\mu_x) = q(n|\mu_{\rm max})$ for all $n \geq m+1$, giving
\begin{equation}\label{eq:tail-methods}
\frac{1}{N}\sum_b \max_x \tr(M_b^\perp\,\rho_x) \leq \frac{T_{m,\mu_{\rm max}}}{N},
\end{equation}
where $T_{m,\mu} := 1 - \Gamma(m+1,\mu)/m!$ is the Poisson tail probability.

For $I>m+1$, the same argument gives a conservative replacement by maximizing each photon-number term separately. Since $q(n|\mu)$ is maximized over $0\leq\mu\leq I$ at $\mu=\min\{n,I\}$,
\begin{equation}\label{eq:tail-general-methods}
\begin{aligned}
T^{\rm wc}_{m,I}
&:=\sum_{n=m+1}^{\infty} q(n|\min\{n,I\})\\
&=\sum_{n=m+1}^{\lfloor I\rfloor}q(n|n)+\sum_{n=\lfloor I\rfloor+1}^{\infty}q(n|I).
\end{aligned}
\end{equation}
The first sum is empty when $\lfloor I\rfloor<m+1$. This reduces to the usual tail $T_{m,I}$ when $I\leq m+1$, and otherwise gives a valid, slightly more conservative, out-of-subspace correction.

\medskip
\noindent\textbf{Bound for trusted scenario.}
When the input intensities $\mu_1 < \cdots < \mu_N$ are known, the subspace contribution is bounded by selecting the $R$ states that are easiest to distinguish:
\begin{equation}\label{eq:opt1-methods}
W_{R,m,\{\mu_i\}} = \frac{1}{N}\!\left(T_{m,\mu_{\rm max}} + \max_{\{\tilde{\mu}_i\}_{i=1}^R \subset \{\mu_i\}_{i=1}^N}\! C_{m,\{\tilde{\mu}_i\}}\right)\!,
\end{equation}
where $C_{m,\{\tilde{\mu}_i\}} := \sum_{n=0}^m \max_i q(n|\tilde{\mu}_i)$. This involves evaluating $\binom{N}{R}$ subsets of states, and can be effectively computed for experimentally relevant parameters.

\medskip
\noindent\textbf{Bound for untrusted scenario.}
When only the intensity bound $\mu_x \leq I$ (with $I \leq m+1$) is assumed, the bound becomes
\begin{equation}\label{eq:opt2-methods}
\widetilde{W}_{R,m,I,N} = \frac{1}{N}\!\left(T_{m,I} + \max_{0 \leq \tilde{\mu}_1,\ldots,\tilde{\mu}_R \leq I} C_{m,\{\tilde{\mu}_i\}}\right)\!.
\end{equation}
For $I>m+1$, we use the same expression with $T_{m,I}$ replaced by the generalized worst-case tail $T^{\rm wc}_{m,I}$ in Eq.~\eqref{eq:tail-general-methods}.
This bound is tight for all cases of interest, i.e. $R \leq N-1$. The bound is also monotonically increasing in $I$: when uncertain about the source calibration, a pessimistic (maximal) estimate of $I$ should be assumed.

\subsection{Numerical calculation of bounds}
\label{sec:methods-numerics}

\begin{table}[h!]
\centering
\begingroup
\footnotesize
\setlength{\tabcolsep}{3pt}
\begin{tabular}{c c c c c c c}
\toprule
$m$  & $N$& $R$ & $W_{R,m,\{\mu_i\}}$ & $\widetilde{W}_{R,m,I,N}$ & $P_{\rm guess}$ & \begin{tabular}{@{}c@{}}Certified genuine\\ resolution\end{tabular} 
\\
\midrule
1 & 2&1& 66.10 & 66.73 & 81.90 & 2/2 \\
2 & 3&2& 66.28 & 67.32 & 65.16 & 2/2 \\
3 & 3&2& 63.45 & 64.97 & 65.16 & 3/3 \\
%4 & 4&2& 52.14 & 52.18 & \color{red} 54.76 & 3/3 \\
4 & 4&3& 56.23 & 57.04 & 54.75 & 3/3 \\
5 & 5&3& 48.67 & 49.64 & 47.98 & 3/3 \\
6 & 6&3& 46.28 & 47.09 & 45.03 & 3/3 \\
7 & 7&3& 40.87 & 41.59 & 40.40 & 3/3 \\
8 & 7&3& 39.65 &  40.86 & 40.40 & 4/3 \\
\bottomrule
\end{tabular}
\endgroup
\caption{\textbf{Certification of genuine resolution.} For each photon-number subspace $m$, we provide the bounds for the witness assuming resolution $R$, for the trusted ($W_{R,m,\{\mu_i\}}$) and untrusted ($\widetilde{W}_{R,m,I,N}$) scenario. Comparing these bounds with experimentally observed score $P_{\rm guess}$, we obtain lower bounds on the certified genuine resolution for each subspace $m$. }
\label{tab:certification}
\end{table}

\noindent\textbf{Trusted scenario.} The bound~\eqref{eq:opt1-methods} requires enumerating $\binom{N}{R}$ subsets and evaluating $C_{m,\{\tilde{\mu}_i\}}$ for each. This is performed by direct computation.

\medskip
\noindent\textbf{Untrusted scenario.}
First, note that the optimization in Eq.~\eqref{eq:opt2-methods} over $R$ real variables $\tilde{\mu}_x$ can be solved heuristically, using standard constrained optimization. Without loss of generality, the intensities are strictly ordered with $\tilde{\mu}_1 = 0$. To enforce monotonic spacing and the global constraint $\tilde{\mu}_R \leq I$, we parametrize the intensities as cumulative sums of non-negative increments and apply a standard optimisation algorithm. While the $\max_i$ operation introduces non-convexity, the objective is well-behaved in practice and converges reliably.

Second, the non-convex continuous optimization can in fact be reformulated as a discrete search with guaranteed global optimality. The key observation is that Poisson distributions are ``well-ordered'': for $\mu < \mu'$, there exists a threshold $n_*$ such that $q(n|\mu) \geq q(n|\mu')$ for $n \leq n_*$ and vice versa. Consequently, the index $i$ maximizing $q(n|\tilde{\mu}_i)$ is an increasing function of $n$, and the integers $\{0,\ldots,m\}$ partition into $R$ contiguous intervals $\mathds{I}_1,\ldots,\mathds{I}_R$ with $\mathds{I}_1 = \{0\}$.

Introducing a formal optimization over such partitions and permuting the order of maximizations yields
\begin{equation}
C^{\rm max}_{R,m,I} = \max_{\{\mathds{I}_i\}} \sum_{i=1}^R \max_{0 \leq \mu_i \leq I} \sum_{n \in \mathds{I}_i} q(n|\mu_i).
\end{equation}
Each inner maximization decouples and admits a closed-form solution. For $i \geq 2$ (with $\mu_1 = 0$), the optimal intensity is
\begin{equation}\label{eq:mu-star}
\mu_i^*(\bm{n}) = \min\!\left\{\exp\!\!\left(\frac{\ln(n_{i+1}-1)! - \ln(n_i - 1)!}{n_{i+1} - n_i}\right),\, I\right\},
\end{equation}
where $n_i$ and $n_{i+1}$ are the endpoints of the interval $\mathds{I}_i$. The outer optimization then reduces to a discrete search over the $\binom{m-1}{R-2}$ partitions of $\{0,\ldots,m\}$ into $R$ contiguous intervals, evaluating an analytical function at each partition.

\subsection{Experimental setup} 
\label{sec:methods-setup}

\begin{figure}[ht!]
    \centering
    \includegraphics[width=0.9\columnwidth]{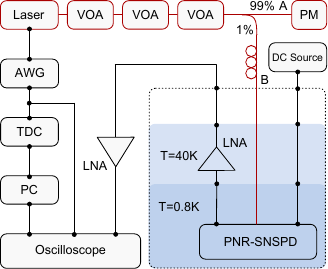}
    \caption{\textbf{Experimental setup for data acquisition.} Coherent states are generated by a 1550 nm pulsed laser driven by an arbitrary waveform generator (AWG). The state intensities are controlled via variable optical attenuators (VOAs) and monitored using a 99:1 beam splitter and a power meter. The attenuated pulses probe a 28-pixel PNR-SNSPD operating at 0.8 K. The detector's output signal, which encodes the number of firing pixels, is amplified both cryogenically (40 K) and at room temperature using low-noise amplifiers (LNAs) before being digitized on a high-resolution oscilloscope. A time-to-digital converter (TDC) measures the total number of laser pulses.}
    \label{fig:figure_setup}
\end{figure}

Figure~\ref{fig:figure_setup} shows the experimental setup used to collect the data required to compute $P_{\rm guess}$ by estimating the probabilities $p(b|\mu_x)$. A pulsed laser is driven by an arbitrary waveform generator (AWG) to produce optical pulses at a wavelength of 1550~nm, with a duration of 33~ps and a repetition rate of 1~MHz. The pulses are attenuated to the desired intensities using three variable optical attenuators (VOAs) and a non-polarizing beam splitter with a 99:1 splitting ratio. The 99\% output (path~A) is monitored with a calibrated power meter, while the 1\% output (path~B) is directed through a polarization controller to a photon-number-resolving (PNR) detector. 

The detector is a PNR-SNSPD array \cite{stasi2024enhanced} with a nominal single-photon detection efficiency of approximately 85\%, a dark-count rate below 100~cps, a timing jitter below 50~ps, and a dynamic range of up to eight photons. The device is operated at 0.8~K in a closed-cycle three-stage cryostat and biased at 95\% of its switching current. The PNR-SNSPD comprises 28 interleaved active nanowires (pixels) connected in parallel with additional on-chip circuitry, enabling simultaneous detection of up to eight photons without electrical crosstalk. In this architecture, as the active nanowires are electrically connected in parallel on-chip, the information on the number of pixels that clicked is encoded in the detector output signals, as illustrated in Fig.~\ref{fig: experiment data analysis}(a). The detector output signals are amplified by a cryogenic amplifier at 40~K, followed by low-noise room-temperature amplification, subsequent digitization, and acquisition using a high-resolution oscilloscope (Tektronix MSO64B, 12-bit ADC, 50~GHz sampling rate). The AWG provides the trigger for the oscilloscope, while a time-to-digital converter (TDC) records the number of laser pulses during data acquisition. 

\medskip
\noindent\textbf{State preparation.}
To prepare the input coherent states, the intensity ratio between the outputs of paths~A and~B is first calibrated with all VOAs set to zero attenuation. For a target input state, the required attenuation—assuming equal attenuation applied by the three VOAs—is estimated from the calibrated intensity ratio and the measured optical power in path~A. Each VOA is subsequently calibrated individually by applying the estimated attenuation to one VOA at a time while setting the remaining two to zero attenuation, and recording the average of multiple power measurements in path~A. The intensity of the input coherent state is then inferred from the experimentally calibrated attenuation values. The overall uncertainty in the calibrated coherent-state intensities is estimated to be approximately 3\%, arising from the combined uncertainties of the power-meter readings, VOA nonlinearity, and beam-splitter ratio calibration.

\medskip
\noindent\textbf{Data acquisition and processing.}
For each input coherent state, $2.5\times10^{5}$ detector output waveforms corresponding to non-zero outcomes are recorded using the oscilloscope, together with time stamps marking the inter-arrival times between consecutive waveforms. The time stamps are post-processed to determine the total acquisition duration, which is used to estimate the total number of laser pulses within the measurement window and the number of zero-click outcomes corresponding to laser pulses for which no detector response is observed. For each recorded waveform, the absolute value of the definite integral is computed over a fixed 2~ns temporal window (shaded region between the dashed lines in the inset of Fig.~\ref{fig: experiment data analysis}(b) using the trapezoidal integration method. Histograms of the integrated areas are then constructed from the $2.5\times10^{5}$ recorded waveforms, as shown in Fig.~\ref{fig: experiment data analysis}(b). The histograms exhibit a series of well-resolved peaks corresponding to different $n$-click outcomes, with clear discrimination thresholds between adjacent peaks (dashed lines in Fig.~\ref{fig: experiment data analysis}(b)). This enables unambiguous assignment of each waveform to a specific photon-number bin, up to eight. The number of events in each bin is counted and normalized by the total number of laser pulses to obtain the click probabilities for each photon-number outcome.

The measured click probability distributions $p(b|\mu_i)$ for the seven input states are reported in Fig~\ref{fig: experiment data analysis}(c).

From these data, the guessing probability $P_{\rm guess} = \frac{1}{N}\sum_{b} \max_{i\in\{1,\ldots,N\}} p(b|\mu_i)$ is computed for each subset of the $N$ weakest input intensities, and compared with the witness bounds derived in Sec.~\ref{sec:methods-witnesses}. The results are given in Table~\ref{tab:certification}. 

From the same guessing probabilities we also extract certified efficiencies relative to the efficiency-limited benchmark model, as described below.

\subsection{Effective efficiency} 
\label{sec:methods-efficiency}

From the experimentally estimated score $P_{\rm guess}$, we can also benchmark the detector against the efficiency-limited PNR detector defined in Eq.~\eqref{eq:eff-lim}, hence leading to a notion of effective efficiency for the detector under test. Specifically, we compute the score obtained by the measurement in Eq.~\eqref{eq:eff-lim}, assuming input dephased coherent states $\rho_{\mu}=\sum_n q(n|\mu)\ketbra{n}$ and an intensity bound $I$ (considering the untrusted scenario). This leads to $p_{\eta,m}(b|\mu):=\tr(M_b^{(\eta,m)}\rho_\mu)$, which allows us to compute the largest guessing probability achievable for the efficiency-limited detector:
\begin{equation}
F_{m,N,I}(\eta)
=
\frac{1}{N}\left[
T_{m,I}
+
\max_{\tilde{\bm{\mu}}\in[0,I]^N}
\right. 
\left.
\sum_{b=0}^{m}
\max_i
p_{\eta,m}(b|\tilde{\mu}_i)
\right].
\label{eq:certified-efficiency}
\end{equation}
where $T_{m,I}$ is the same conservative tail-correction as in Eq.~\eqref{eq:opt2-methods}.  For $I>m+1$, it is replaced by $T^{\rm wc}_{m,I}$ in Eq.~\eqref{eq:tail-general-methods}.

Finally, by equating the observed score $P_{\rm guess}$ with Eq. \eqref{eq:certified-efficiency}, i.e. setting 
\begin{equation}
F_{m,N,I}(\eta_m^*) = P_{\rm guess},
\end{equation}
we get the effective efficiency $\eta_m^*$ for the PNR detector under test, in the subspace with up to $m$ photons. Operationally, this means that the observed data is not compatible with an efficiency-limited detector with efficiency $\eta < \eta_m^*$. Hence the detector under test cannot be described by an efficiency-limited detector with $\eta < \eta_m^*$ followed by classical post-processing. Note that, for the trusted scenario, the score is evaluated as in Eq. \eqref{eq:certified-efficiency}, but assuming calibrated intensities $\{\mu_i\}_{i=1}^N$.

\begin{table}[h]
\centering
\begingroup
\footnotesize
\setlength{\tabcolsep}{5pt}
\begin{tabular}{c c c c c}
\toprule
$m$ & $N$ & $I_{}$ & trusted $\eta_m^*$ & untrusted $\eta_m^*$ \\
\midrule
1 & 2 & 1.1921 & $86.89\%$ & $82.49\%$ \\
2 & 3 & 2.4088 & $82.54\%$ & $78.22\%$ \\
3 & 3 & 2.4088 & $84.20\%$ & $81.59\%$ \\
4 & 4 & 3.5817 & $82.71\%$ & $80.00\%$ \\
5 & 5 & 4.7928 & $82.20\%$ & $79.29\%$ \\
6 & 6 & 7.1993 & $77.07\%$ & $72.04\%$ \\
7 & 7 & 8.2133 & $77.42\%$ & $72.38\%$ \\
8 & 7 & 8.2133 & $78.47\%$ & $74.27\%$ \\
\bottomrule
\end{tabular}
\endgroup
\caption{\textbf{Effective efficiencies.} For each photon-number subspace with up to $m$ photons, the table reports the certified effective efficiency $\eta_m^*$ for our PNR detector, based on the experimentally estimated score $P_{\rm guess}$. Both the trusted and untrusted scenarios are considered; for the latter, the intensity bound is taken as $I=1.03\max_{i\leq N}\mu_i$.}
\label{tab:certified-efficiencies}
\end{table}

\begin{acknowledgments}
The numerical data, analysis code, and an interactive web application for evaluating the resolution witnesses are available at \url{https://github.com/jefpauwels/resolution_certification}. We thank Armin Tavakoli for early discussions. We acknowledge support from the Swiss State Secretariat for Research and Innovation (SERI) [contract number UeM019-3]; the Swiss National Science Foundation (SNSF) under the Korean-Swiss Science and Technology Programme [grant number 10.005.736]; National Research Council Canada (NRC) Collaborative Science, Technology and Innovation program (CSTIP) [grant number QSP043]; Swedish Research Council [grant number 2024-05341]; and the Wallenberg Initiative on Networks and Quantum Information (WINQ). The authors declare no competing interests.
\end{acknowledgments}

\clearpage

\bibliography{refsv2}

@ARTICLE{taylor2023,
  title   = "{Low-noise single-photon counting superconducting nanowire
             detectors at infrared wavelengths up to 29 µm}",
  author  = "Taylor, Gregor G and Walter, Alexander B and Korzh, Boris and
             Bumble, Bruce and Patel, Sahil R and Allmaras, Jason P and Beyer,
             Andrew D and O'Brient, Roger and Shaw, Matthew D and Wollman, Emma
             E",
  journal = "Optica",
  volume  =  10,
  pages   =  1672,
  year    =  2023,
  url     = "http://dx.doi.org/10.1364/OPTICA.509337"
}

@ARTICLE{resta2023,
  title   = "{Gigahertz Detection Rates and Dynamic Photon-Number Resolution
             with Superconducting Nanowire Arrays}",
  author  = "Resta, Giovanni V and Stasi, Lorenzo and Perrenoud, Matthieu and
             El-Khoury, Sylvain and Brydges, Tiff and Thew, Rob and Zbinden,
             Hugo and Bussières, Félix",
  journal = "Nano Lett.",
  volume  =  23,
  number  =  13,
  pages   = "6018-6026",
  year    =  2023,
  url     = "http://dx.doi.org/10.1021/acs.nanolett.3c01228"
}

@ARTICLE{korzh2020,
  title     = "{Demonstration of sub-3 ps temporal resolution with a
               superconducting nanowire single-photon detector}",
  author    = "Korzh, Boris and Zhao, Qing-Yuan and Allmaras, Jason P and
               Frasca, Simone and Autry, Travis M and Bersin, Eric A and Beyer,
               Andrew D and Briggs, Ryan M and Bumble, Bruce and Colangelo,
               Marco and Crouch, Garrison M and Dane, Andrew E and Gerrits,
               Thomas and Lita, Adriana E and Marsili, Francesco and Moody,
               Galan and Peña, Cristián and Ramirez, Edward and Rezac, Jake D
               and Sinclair, Neil and Stevens, Martin J and Velasco, Angel E and
               Verma, Varun B and Wollman, Emma E and Xie, Si and Zhu, Di and
               Hale, Paul D and Spiropulu, Maria and Silverman, Kevin L and
               Mirin, Richard P and Nam, Sae Woo and Kozorezov, Alexander G and
               Shaw, Matthew D and Berggren, Karl K",
  journal   = "Nat. Photonics",
  publisher = "Nature Publishing Group",
  volume    =  14,
  number    =  4,
  pages     = "250-255",
  year      =  2020,
  url       = "https://www.nature.com/articles/s41566-020-0589-x"
}

@ARTICLE{oripov2023,
  title     = "{A superconducting nanowire single-photon camera with 400,000
               pixels}",
  author    = "Oripov, B G and Rampini, D S and Allmaras, J and Shaw, M D and
               Nam, S W and Korzh, B and McCaughan, A N",
  journal   = "Nature",
  publisher = "Nature Publishing Group",
  volume    =  622,
  number    =  7984,
  pages     = "730-734",
  year      =  2023,
  url       = "https://www.nature.com/articles/s41586-023-06550-2"
}

@ARTICLE{tenne2019,
  title     = "{Super-resolution enhancement by quantum image scanning
               microscopy}",
  author    = "Tenne, Ron and Rossman, Uri and Rephael, Batel and Israel,
               Yonatan and Krupinski-Ptaszek, Alexander and Lapkiewicz, Radek
               and Silberberg, Yaron and Oron, Dan",
  journal   = "Nat. Photonics",
  publisher = "Springer Science and Business Media LLC",
  volume    =  13,
  number    =  2,
  pages     = "116-122",
  year      =  2019,
  doi       = "10.1038/s41566-018-0324-z"
}

@ARTICLE{natarajan2013,
  title   = "{Quantum detector tomography of a time-multiplexed superconducting
             nanowire single-photon detector at telecom wavelengths}",
  author  = "Natarajan, Chandra M and Zhang, Lijian and Coldenstrodt-Ronge,
             Hendrik and Donati, Gaia and Dorenbos, Sander N and Zwiller, Val
             and Walmsley, Ian A and Hadfield, Robert H",
  journal = "Opt. Express",
  volume  =  21,
  number  =  1,
  pages   = "893-902",
  year    =  2013,
  doi     = "10.1364/OE.21.000893"
}

@ARTICLE{fukuda2011,
  title     = "{Titanium-based transition-edge photon number resolving detector
               with 98\% detection efficiency with index-matched small-gap fiber
               coupling}",
  author    = "Fukuda, Daiji and Fujii, Go and Numata, Takayuki and Amemiya,
               Kuniaki and Yoshizawa, Akio and Tsuchida, Hidemi and Fujino,
               Hidetoshi and Ishii, Hiroyuki and Itatani, Taro and Inoue,
               Shuichiro and Zama, Tatsuya",
  journal   = "Opt. Express",
  publisher = "Optica Publishing Group",
  volume    =  19,
  number    =  2,
  pages     = "870-875",
  year      =  2011,
  doi       = "10.1364/OE.19.000870"
}

@ARTICLE{afek2010,
  title     = "{High-NOON states by mixing quantum and classical light}",
  author    = "Afek, Itai and Ambar, Oron and Silberberg, Yaron",
  journal   = "Science",
  publisher = "American Association for the Advancement of Science (AAAS)",
  volume    =  328,
  number    =  5980,
  pages     = "879-881",
  year      =  2010,
  doi       = "10.1126/science.1188172"
}

@ARTICLE{moroder2009,
  title     = "{Detector decoy quantum key distribution}",
  author    = "Moroder, Tobias and Curty, Marcos and Lütkenhaus, Norbert",
  journal   = "New J. Phys.",
  publisher = "IOP Publishing",
  volume    =  11,
  number    =  4,
  pages     =  045008,
  year      =  2009,
  doi       = "10.1088/1367-2630/11/4/045008"
}

@ARTICLE{eaton2023,
  title     = "{Resolution of 100 photons and quantum generation of unbiased
               random numbers}",
  author    = "Eaton, Miller and Hossameldin, Amr and Birrittella, Richard J and
               Alsing, Paul M and Gerry, Christopher C and Dong, Hai and Cuevas,
               Chris and Pfister, Olivier",
  journal   = "Nat. Photonics",
  publisher = "Springer Science and Business Media LLC",
  volume    =  17,
  number    =  1,
  pages     = "106-111",
  year      =  2023,
  doi       = "10.1038/s41566-022-01105-9"
}

@ARTICLE{you2021,
  title     = "{Scalable multiphoton quantum metrology with neither pre- nor
               post-selected measurements}",
  author    = "You, Chenglong and Hong, Mingyuan and Bierhorst, Peter and Lita,
               Adriana E and Glancy, Scott and Kolthammer, Steve and Knill,
               Emanuel and Nam, Sae Woo and Mirin, Richard P and Magaña-Loaiza,
               Omar S and Gerrits, Thomas",
  journal   = "Appl. Phys. Rev.",
  publisher = "AIP Publishing",
  volume    =  8,
  number    =  4,
  pages     =  041406,
  year      =  2021,
  doi       = "10.1063/5.0063294"
}

@ARTICLE{winnel2024,
  title     = "{Deterministic preparation of optical squeezed cat and
               Gottesman-Kitaev-Preskill states}",
  author    = "Winnel, Matthew S and Guanzon, Joshua J and Singh, Deepesh and
               Ralph, Timothy C",
  journal   = "Phys. Rev. Lett.",
  publisher = "American Physical Society (APS)",
  volume    =  132,
  number    =  23,
  pages     =  230602,
  year      =  2024,
  doi       = "10.1103/PhysRevLett.132.230602"
}

@ARTICLE{kaneda2019,
  title   = "{High-efficiency single-photon generation via large-scale active
             time multiplexing}",
  author  = "Kaneda, F and Kwiat, P G",
  journal = "Sci. Adv.",
  volume  =  5,
  number  =  10,
  pages   = "eaaw8586",
  year    =  2019,
  doi     = "10.1126/sciadv.aaw8586"
}

@ARTICLE{konno2024,
  title     = "{Logical states for fault-tolerant quantum computation with
               propagating light}",
  author    = "Konno, Shunya and Asavanant, Warit and Hanamura, Fumiya and
               Nagayoshi, Hironari and Fukui, Kosuke and Sakaguchi, Atsushi and
               Ide, Ryuhoh and China, Fumihiro and Yabuno, Masahiro and Miki,
               Shigehito and Terai, Hirotaka and Takase, Kan and Endo, Mamoru
               and Marek, Petr and Filip, Radim and van Loock, Peter and
               Furusawa, Akira",
  journal   = "Science",
  publisher = "American Association for the Advancement of Science",
  volume    =  383,
  number    =  6680,
  pages     = "289-293",
  year      =  2024,
  doi       = "10.1126/science.adk7560"
}

@ARTICLE{becerra2015,
  title     = "{Photon number resolution enables quantum receiver for realistic
               coherent optical communications}",
  author    = "Becerra, F E and Fan, J and Migdall, A",
  journal   = "Nat. Photonics",
  publisher = "Springer Science and Business Media LLC",
  volume    =  9,
  number    =  1,
  pages     = "48-53",
  year      =  2015,
  doi       = "10.1038/nphoton.2014.280"
}

@ARTICLE{humphreys2015,
  title     = "{Tomography of photon-number resolving continuous-output
               detectors}",
  author    = "Humphreys, Peter C and Metcalf, Benjamin J and Gerrits, Thomas
               and Hiemstra, Thomas and Lita, Adriana E and Nunn, Joshua and
               Nam, Sae Woo and Datta, Animesh and Kolthammer, W Steven and
               Walmsley, Ian A",
  journal   = "New J. Phys.",
  publisher = "IOP Publishing",
  volume    =  17,
  number    =  10,
  pages     =  103044,
  year      =  2015,
  doi       = "10.1088/1367-2630/17/10/103044"
}

@ARTICLE{ding2025,
  title     = "{Photon-number-resolving single-photon detector with a system
               detection efficiency of 98\% and photon-number resolution of 32}",
  author    = "Ding, Chaomeng and Zhang, Xingyu and Xiong, Jiamin and Xiao, You
               and Zhang, Tianzhu and Huang, Jia and Xu, Hongxin and Liu, Xiaoyu
               and You, Lixing and Wang, Zhen and Li, Hao",
  journal   = "ACS Photonics",
  publisher = "American Chemical Society (ACS)",
  volume    =  12,
  number    =  9,
  pages     = "4924-4931",
  year      =  2025,
  doi       = "10.1021/acsphotonics.5c00508"
}

@ARTICLE{dynes2018,
  title     = "{Testing the photon-number statistics of a quantum key
               distribution light source}",
  author    = "Dynes, J F and Lucamarini, M and Patel, K A and Sharpe, A W and
               Ward, M B and Yuan, Z L and Shields, A J",
  journal   = "Opt. Express",
  publisher = "Optica Publishing Group",
  volume    =  26,
  number    =  18,
  pages     = "22733-22749",
  year      =  2018,
  doi       = "10.1364/OE.26.022733"
}

@ARTICLE{davis2022,
  title     = "{Improved heralded single-photon source with a
               photon-number-resolving superconducting nanowire detector}",
  author    = "Davis, Samantha I and Mueller, Andrew and Valivarthi, Raju and
               Lauk, Nikolai and Narvaez, Lautaro and Korzh, Boris and Beyer,
               Andrew D and Cerri, Olmo and Colangelo, Marco and Berggren,
               Karl K and Shaw, Matthew D and Xie, Si and Sinclair, Neil and
               Spiropulu, Maria",
  journal   = "Phys. Rev. Applied",
  publisher = "APS",
  volume    =  18,
  number    =  6,
  pages     =  064007,
  year      =  2022,
  doi       = "10.1103/PhysRevApplied.18.064007"
}

@ARTICLE{kok2007,
  title     = "{Linear optical quantum computing with photonic qubits}",
  author    = "Kok, Pieter and Munro, W J and Nemoto, Kae and Ralph, T C and
               Dowling, Jonathan P and Milburn, G J",
  journal   = "Rev. Mod. Phys.",
  publisher = "American Physical Society",
  volume    =  79,
  number    =  1,
  pages     = "135-174",
  year      =  2007,
  doi       = "10.1103/RevModPhys.79.135"
}

@ARTICLE{cahall2017,
  title   = "{Multi-photon detection using a conventional superconducting
             nanowire single-photon detector}",
  author  = "Cahall, Clinton and Nicolich, Kathryn L and Islam, Nurul T and
             Lafyatis, Gregory P and Miller, Aaron J and Gauthier, Daniel J and
             Kim, Jungsang",
  journal = "Optica",
  volume  =  4,
  number  =  12,
  pages   = "1534-1535",
  year    =  2017,
  doi     = "10.1364/OPTICA.4.001534"
}

@article{Fiurasek2001,
  title = {Maximum-likelihood estimation of quantum measurement},
  author = {Fiur\'a\ifmmode \check{s}\else \v{s}\fi{}ek, Jarom\'{\i}r},
  journal = {Phys. Rev. A},
  volume = {64},
  issue = {2},
  pages = {024102},
  numpages = {4},
  year = {2001},
  month = {Jul},
  publisher = {American Physical Society},
  doi = {10.1103/PhysRevA.64.024102}
}

@article{Luis1999,
  title = {Complete Characterization of Arbitrary Quantum Measurement Processes},
  author = {Luis, A. and S\'anchez-Soto, L. L.},
  journal = {Phys. Rev. Lett.},
  volume = {83},
  issue = {18},
  pages = {3573--3576},
  numpages = {4},
  year = {1999},
  month = {Nov},
  publisher = {American Physical Society},
  doi = {10.1103/PhysRevLett.83.3573}
}

@article{Dariano2004,
  title = {Quantum Calibration of Measurement Instrumentation},
  author = {D'Ariano, Giacomo Mauro and Maccone, Lorenzo and {Lo Presti}, Paoloplacido},
  journal = {Phys. Rev. Lett.},
  volume = {93},
  issue = {25},
  pages = {250407},
  numpages = {4},
  year = {2004},
  month = {Dec},
  publisher = {American Physical Society},
  doi = {10.1103/PhysRevLett.93.250407}
}

@article{Uola2023,
  title = {Colloquium: Incompatible measurements in quantum information science},
  author = {G\"uhne, Otfried and Haapasalo, Erkka and Kraft, Tristan and Pellonp\"a\"a, Juha-Pekka and Uola, Roope},
  journal = {Rev. Mod. Phys.},
  volume = {95},
  issue = {1},
  pages = {011003},
  numpages = {25},
  year = {2023},
  month = {Feb},
  publisher = {American Physical Society},
  doi = {10.1103/RevModPhys.95.011003}
}

@article{oszmaniec2017simulating,
  title={Simulating positive-operator-valued measures with projective measurements},
  author={Oszmaniec, Micha{\l} and Guerini, Leonardo and Wittek, Peter and Ac{\'\i}n, Antonio},
  journal={Phys. Rev. Lett.},
  volume={119},
  number={19},
  pages={190501},
  year={2017},
  publisher={American Physical Society},
  doi={10.1103/PhysRevLett.119.190501}
}

@article{ioannou2022simulability,
  title={Simulability of High-Dimensional Quantum Measurements},
  author={Ioannou, Marie and Sekatski, Pavel and Designolle, S{\'e}bastien and Jones, Benjamin D. M. and Uola, Roope and Brunner, Nicolas},
  journal={Phys. Rev. Lett.},
  volume={129},
  number={19},
  pages={190401},
  year={2022},
  publisher={American Physical Society},
  doi={10.1103/PhysRevLett.129.190401}
}

@article{jonsson2019evaluating,
  title={Evaluating the performance of photon-number-resolving detectors},
  author={J{\"o}nsson, Mattias and Bj{\"o}rk, Gunnar},
  journal={Phys. Rev. A},
  volume={99},
  number={4},
  pages={043822},
  year={2019},
  publisher={APS},
  doi={10.1103/PhysRevA.99.043822}
}

@article{provaznik2020benchmarking,
  title={Benchmarking photon number resolving detectors},
  author={Provazn{\'\i}k, Jan and Lachman, Luk{\'a}{\v{s}} and Filip, Radim and Marek, Petr},
  journal={Opt. Express},
  volume={28},
  number={10},
  pages={14839--14849},
  year={2020},
  publisher={Optica Publishing Group},
  doi={10.1364/OE.389619}
}

@article{guerini2017operational,
  title={Operational framework for quantum measurement simulability},
  author={Guerini, Leonardo and Bavaresco, Jessica and Terra Cunha, Marcelo and Ac{\'\i}n, Antonio},
  journal={J. Math. Phys.},
  volume={58},
  number={9},
  pages={092102},
  year={2017},
  publisher={AIP Publishing},
  doi={10.1063/1.4994303}
}

@article{d2005classical,
  title={Classical randomness in quantum measurements},
  author={D'Ariano, Giacomo Mauro and {Lo Presti}, Paoloplacido and Perinotti, Paolo},
  journal={J. Phys. A: Math. Gen.},
  volume={38},
  number={26},
  pages={5979--5991},
  year={2005},
  publisher={IOP Publishing},
  doi={10.1088/0305-4470/38/26/010}
}

@Article{Craiciu2023,
  author    = {Craiciu, Ioana and Korzh, Boris and Beyer, Andrew D. and Mueller, Andrew and Allmaras, Jason P. and Narváez, Lautaro and Spiropulu, Maria and Bumble, Bruce and Lehner, Thomas and Wollman, Emma E. and Shaw, Matthew D.},
  journal   = {Optica},
  title     = {High-speed detection of 1550nm single photons with superconducting nanowire detectors},
  year      = {2023},
  issn      = {2334-2536},
  month     = jan,
  number    = {2},
  pages     = {183},
  volume    = {10},
  doi       = {10.1364/optica.478960},
  publisher = {Optica Publishing Group},
}

@Misc{Gill2002,
  author        = {Gill, Richard D.},
  title         = {Time, finite statistics, and {Bell's} fifth position},
  howpublished  = {In \textit{Foundations of Probability and Physics II} (ed. Khrennikov, A.), Vol.~5 of \textit{Mathematical Modelling in Physics, Engineering and Cognitive Sciences}, 179--206 (V{\"a}xj{\"o} University Press, V{\"a}xj{\"o}, 2003)},
}

@article{Sanguinetti2014,
  title = {Quantum Random Number Generation on a Mobile Phone},
  author = {Sanguinetti, Bruno and Martin, Anthony and Zbinden, Hugo and Gisin, Nicolas},
  journal = {Phys. Rev. X},
  volume = {4},
  issue = {3},
  pages = {031056},
  numpages = {6},
  year = {2014},
  month = {Sep},
  publisher = {American Physical Society},
  doi = {10.1103/PhysRevX.4.031056}
}

@Article{VanHimbeeck2017,
  author    = {Van Himbeeck, Thomas and Woodhead, Erik and Cerf, Nicolas J. and Garc{\'{i}}a-Patr{\'{o}}n, Ra{\'{u}}l and Pironio, Stefano},
  journal   = {{Quantum}},
  title     = {Semi-device-independent framework based on natural physical assumptions},
  year      = {2017},
  issn      = {2521-327X},
  month     = nov,
  pages     = {33},
  volume    = {1},
  doi       = {10.22331/q-2017-11-18-33},
  publisher = {{Verein zur F{\''{o}}rderung des Open Access Publizierens in den Quantenwissenschaften}},
}

@article{Knill2001,
  author = {Knill, E. and Laflamme, R. and Milburn, G. J.},
  title = {A scheme for efficient quantum computation with linear optics},
  journal = {Nature},
  volume = {409},
  pages = {46--52},
  year = {2001},
  doi = {10.1038/35051009}
}

@article{Bartolucci2023,
  author = {Bartolucci, Sara and Birchall, Patrick and Bomb{\'\i}n, H{\'e}ctor and Cable, Hugo and Dawson, Chris and Gimeno-Segovia, Mercedes and Johnston, Eric and Kieling, Konrad and Nickerson, Naomi and Pant, Mihir and Pastawski, Fernando and Rudolph, Terry and Sparrow, Chris},
  title = {Fusion-based quantum computation},
  journal = {Nat. Commun.},
  volume = {14},
  pages = {912},
  year = {2023},
  doi = {10.1038/s41467-023-36493-1}
}

@article{Zhong2020,
  author = {Zhong, Han-Sen and Wang, Hui and Deng, Yu-Hao and Chen, Ming-Cheng and Peng, Li-Chao and Luo, Yi-Han and Qin, Jian and Wu, Dian and Ding, Xing and Hu, Yi and Hu, Peng and Yang, Xiao-Yan and Zhang, Wei-Jun and Li, Hao and Li, Yuxuan and Jiang, Xiao and Gan, Lin and Yang, Guangwen and You, Lixing and Wang, Zhen and Li, Li and Liu, Nai-Le and Lu, Chao-Yang and Pan, Jian-Wei},
  title = {Quantum computational advantage using photons},
  journal = {Science},
  volume = {370},
  number = {6523},
  pages = {1460--1463},
  year = {2020},
  doi = {10.1126/science.abe8770}
}

@article{Madsen2022,
  author = {Madsen, Lars S. and Laudenbach, Fabian and Askarani, Mohsen Falamarzi and Rortais, Fabien and Vincent, Trevor and Bulmer, Jacob F. F. and Miatto, Filippo M. and Neuhaus, Leonhard and Helt, Lukas G. and Collins, Matthew J. and Lita, Adriana E. and Gerrits, Thomas and Nam, Sae Woo and Vaidya, Varun D. and Menotti, Matteo and Dhand, Ish and Vernon, Zachary and Quesada, Nicol{\'a}s and Lavoie, Jonathan},
  title = {Quantum computational advantage with a programmable photonic processor},
  journal = {Nature},
  volume = {606},
  number = {7912},
  pages = {75--81},
  year = {2022},
  doi = {10.1038/s41586-022-04725-x}
}

@article{Lita2008,
  author = {Lita, Adriana E. and Miller, Aaron J. and Nam, Sae Woo},
  title = {Counting near-infrared single-photons with 95\% efficiency},
  journal = {Opt. Express},
  volume = {16},
  number = {5},
  pages = {3032--3040},
  year = {2008},
  doi = {10.1364/OE.16.003032}
}

@article{Divochiy2008,
  author = {Divochiy, A. and Marsili, F. and Bitauld, D. and Gaggero, A. and Leoni, R. and Mattioli, F. and Korneev, A. and Seleznev, V. and Kaurova, N. and Minaeva, O. and Gol'tsman, G. and Lagoudakis, K. G. and Benkhaoul, M. and L{\'e}vy, F. and Fiore, A.},
  title = {Superconducting nanowire photon-number-resolving detector at telecommunication wavelengths},
  journal = {Nat. Photonics},
  volume = {2},
  number = {5},
  pages = {302--306},
  year = {2008},
  doi = {10.1038/nphoton.2008.51}
}

@article{Lundeen2009,
  author = {Lundeen, Jeff S. and Feito, Alejandra and Coldenstrodt-Ronge, Hendrik and Pregnell, Kevin L. and Silberhorn, Christine and Ralph, Timothy C. and Eisert, Jens and Plenio, Martin B. and Walmsley, Ian A.},
  title = {Tomography of quantum detectors},
  journal = {Nat. Phys.},
  volume = {5},
  pages = {27--30},
  year = {2009},
  doi = {10.1038/nphys1133}
}

@article{stasi2024enhanced,
  title={Enhanced detection rate and high photon-number efficiencies with a scalable parallel {SNSPD}},
  author={Stasi, Lorenzo and Taher, Towsif and Resta, Giovanni V and Zbinden, Hugo and Thew, Rob and Bussi{\`e}res, F{\'e}lix},
  journal={ACS Photonics},
  volume={12},
  number={1},
  pages={320--329},
  year={2025},
  publisher={ACS Publications},
  doi={10.1021/acsphotonics.4c01680}
}

@article{cheng2023100,
  title={A 100-pixel photon-number-resolving detector unveiling photon statistics},
  author={Cheng, Risheng and Zhou, Yiyu and Wang, Sihao and Shen, Mohan and Taher, Towsif and Tang, Hong X},
  journal={Nat. Photonics},
  volume={17},
  number={1},
  pages={112--119},
  year={2023},
  publisher={Nature Publishing Group UK London},
  doi={10.1038/s41566-022-01119-3}
}

@misc{brask2026,
      title={Quantum correlations in prepare-and-measure scenarios and their semi-device-independent applications}, 
      author={Jonatan Bohr Brask and Nicolas Brunner and Jef Pauwels and Davide Rusca and Armin Tavakoli},
      year={2026},
      howpublished={Preprint at \url{https://arXiv.org/abs/2603.23604}}
}

@article{zhu2020resolving,
  title={Resolving photon numbers using a superconducting nanowire with impedance-matching taper},
  author={Zhu, Di and Colangelo, Marco and Chen, Changchen and Korzh, Boris A and Wong, Franco NC and Shaw, Matthew D and Berggren, Karl K},
  journal={Nano Lett.},
  volume={20},
  number={5},
  pages={3858--3863},
  year={2020},
  publisher={ACS Publications},
  doi={10.1021/acs.nanolett.0c00985}
}

@article{los2024high,
  title={High-performance photon number resolving detectors for 850--950 nm wavelength range},
  author={Los, J. W. Niels and Sidorova, Mariia and Lopez-Rodriguez, Bruno and Qualm, Patrick and Chang, Jin and Steinhauer, Stephan and Zwiller, Val and Zadeh, Iman Esmaeil},
  journal={APL Photonics},
  volume={9},
  number={6},
  pages={066101},
  year={2024},
  publisher={AIP Publishing},
  doi={10.1063/5.0204340}
}

@article{psiquantum2025manufacturable,
  title={A manufacturable platform for photonic quantum computing},
  author={{PsiQuantum team}},
  journal={Nature},
  volume={641},
  number={8064},
  pages={876--883},
  year={2025},
  publisher={Nature Publishing Group UK London},
  doi={10.1038/s41586-025-08820-7}
}

@article{morais2024precisely,
  title={Precisely determining photon-number in real time},
  author={Morais, Leonardo Assis and Weinhold, Till and de Almeida, Marcelo Pereira and Combes, Joshua and Rambach, Markus and Lita, Adriana and Gerrits, Thomas and Nam, Sae Woo and White, Andrew G and Gillett, Geoff},
  journal={Quantum},
  volume={8},
  pages={1355},
  year={2024},
  publisher={Verein zur F{\"o}rderung des Open Access Publizierens in den Quantenwissenschaften},
  doi={10.22331/q-2024-05-23-1355}
}

\appendix

\section{Statistical analysis}
\label{sec:methods-stats}

The estimator of $P_{\rm guess}$ is a uniformly weighted average over the $N$ input states. For the statistical analysis we use $n_0=2.5\times10^5$ recorded events for each non-vacuum intensity. 
The vacuum input is treated as exact, consistent with the data model $p(0|0)=1$ and with dark count rates of the order of $10^{-9}$.

Applying the Azuma--Hoeffding inequality~\cite{Gill2002}, to the estimator for any $\delta > 0$:
\begin{align}
p_{\rm H} &:=\Pr\!\left(P_{\rm guess} \geq W + \delta\right) \leq \exp(-2n_{\rm eff}\,\delta^2)\,, \\
n_{\rm eff}
&:=\frac{N^2}{(N-1)/n_0},
\end{align}
where $W$ is the applicable witness bound. We obtain the corresponding Gaussian-equivalent one-sided significances $Z=\Phi^{-1}(1-p_{\rm H})$, where $\Phi$ is the standard normal cumulative distribution function; since $p_{\rm H}$ is an upper bound, these are conservative lower bounds on $Z$.

Table~\ref{tab:stats-significance} reports the statistical analysis for the three main certification rows.

\begin{table*}
\centering
\begingroup
\footnotesize
\setlength{\tabcolsep}{4pt}
\begin{tabular}{c c c c c c c}
\toprule
$(m,N,R)$ & source model & $I$ & bound [\%] & $\delta$ [\%] & $p_{\rm H}$ & $Z$ \\
\midrule
$(1,2,1)$ & trusted & -- & 66.10 & 15.80 & $1.39\times 10^{-21694}$ & $316.1\sigma$ \\
$(1,2,1)$ & untrusted & 1.1921 & 66.73 & 15.17 & $1.51\times 10^{-19999}$ & $303.5\sigma$ \\
$(3,3,2)$ & trusted & -- & 63.45 & 1.71 & $2.76\times 10^{-286}$ & $36.1\sigma$ \\
$(3,3,2)$ & untrusted & 2.4088 & 64.97 & 0.189 & $3.14\times 10^{-4}$ & $3.42\sigma$ \\
$(8,7,3)$ & trusted & -- & 39.65 & 0.752 & $5.94\times 10^{-101}$ & $21.3\sigma$ \\
$(8,7,3)$ & untrusted & 8.2133 & 40.86 & -0.462 & 1 & no violation \\
\bottomrule
\end{tabular}
\endgroup
\caption{\textbf{Statistical significances.} Here $\delta=P_{\rm guess}-B$, where $B$ is either the trusted witness $W_{R,m,\{\mu_i\}}$ or the intensity-bounded witness $\widetilde{W}_{R,m,I,N}$. The intensity rows use $I=1.03\max_i\mu_i$. The effective counts are $n_{\rm eff}=1.00\times10^6$, $1.125\times10^6$, and $2.042\times10^6$ for the $N=2$, $N=3$, and $N=7$ rows, respectively.}
\label{tab:stats-significance}
\end{table*}

\end{document}